\documentclass[10pt,twocolumn]{IEEEtran}
\IEEEoverridecommandlockouts

\usepackage{color}
\usepackage{graphicx}
\usepackage{amsmath}
\usepackage{amssymb}
\usepackage{algorithmic}
\usepackage{amsmath}
\usepackage{multirow}
\usepackage{booktabs}
\usepackage{array}
\usepackage{amsthm}
\usepackage{stfloats}
\usepackage{caption}
\usepackage{subfigure}
\usepackage{bm}
\usepackage{booktabs}
\usepackage{setspace}
\usepackage{epstopdf}
\usepackage{makecell}
\usepackage{multicol}
\usepackage[linesnumbered,ruled]{algorithm2e}
\usepackage{amsmath,siunitx}
\usepackage{mathtools}
\usepackage{multicol}

\allowdisplaybreaks[4]

\newcommand{\be}{\begin{equation}}
\newcommand{\ee}{\end{equation}}
\newcommand{\bea}{\begin{eqnarray}}
\newcommand{\eea}{\end{eqnarray}}
\newcommand{\ba}{\begin{array}}
\newcommand{\ea}{\end{array}}

\newcommand{\PreserveBackslash}[1]{\let\temp=\\#1\let\\=\temp}

\newcolumntype{C}[1]{>{\PreserveBackslash\centering}p{#1}}

\newcolumntype{R}[1]{>{\PreserveBackslash\raggedleft}p{#1}}

\newcolumntype{L}[1]{>{\PreserveBackslash\raggedright}p{#1}}

\title{
Two-timescale Optimization for Hybrid Mechanically and Electronically Tunable 6DMA Aided Communication }
\vspace{-0.6cm}
\author{Yuyan Zhou, \,\thanks{Y. Zhou and H. Hua are with the School of Science and Engineering, The Chinese University of Hong Kong (Shenzhen), Guangdong 518172, China (email: yuyanzhou1@link.cuhk.edu.cn, huahaocheng@cuhk.edu.cn).} Haocheng Hua, \emph{Member, IEEE,}\, Jie Xu, \emph{Fellow, IEEE,} \thanks{J. Xu is with the School of Science and Engineering, the Shenzhen Future Network of Intelligence Institute (FNii-Shenzhen), and the Guangdong Provincial Key Laboratory of Future Networks of Intelligence, The Chinese University of Hong Kong (Shenzhen), Guandong 518172, China (email: xujie@cuhk.edu.cn).}and Rui Zhang, \emph{Fellow, IEEE} \thanks{R. Zhang is with the Department of Electrical and Computer Engineering, National University of Singapore, Singapore 117583 (e-mail: elezhang@nus.edu.sg).} \vspace{-0.95cm}}
\begin{document}
\maketitle
\pagestyle{empty}
\thispagestyle{empty}

\begin{abstract}
This letter proposes a hybrid mechanically and electronically tunable six-dimensional movable antenna (6DMA) base station (BS) architecture for future wireless communication networks. Such BS consists of multiple antenna arrays that are mechanically movable along a circular rail to adapt to the horizontal user hotspots, and each array is equipped with pattern reconfigurable antennas (PRAs) that are capable of electronically switching among a set of specified beam patterns to cater to the instantaneous user channels. The mechanical adjustment provides wide-angle coverage but suffers from slow response, while the electronic tuning enables rapid beam reconfiguration but with limited angular range.
To effectively combine their complementary advantages, we propose to jointly design both mechanical and electronic configurations to maximize the average sum-rate of users via a two-timescale optimization approach, in which the array positions are optimized on the long timescale according to large-scale user distribution statistics, and the pattern selection vectors are optimized on the short timescale to enable fast beam alignment based on the instantaneous user locations. An alternating optimization algorithm based on the Monte Carlo sampling method is developed to solve the problem efficiently. Finally, simulation results show that our proposed design achieves significant performance gains over various benchmark schemes.
\end{abstract}

\vspace{-0.5cm}

\maketitle

\vspace{-0.3cm}
\section{Introduction}
Multiple-input multiple-output (MIMO) technology has been a cornerstone of modern wireless communication systems, offering diversity gains and spatial multiplexing gains. To meet stringent requirements of future sixth-generation (6G) wireless networks, such as unprecedented high data rate and ultra-low latency \cite{6G}, various advanced MIMO architectures have been developed, including extremely large-scale MIMO (XL-MIMO) for finer spatial resolution \cite{XL-MIMO} and cell-free massive MIMO for distributed MIMO cooperation \cite{cell-free}.

However, the aforementioned MIMO systems employ fixed-position antennas (FPAs), which cannot fully adapt to spatial and temporal channel variations. To address this limitation, movable antenna (MA) has been introduced \cite{MA1} to exploit spatial channel variations for enhanced performance via adjusting antenna positions \cite{MA2}. More recently, six-dimensional movable antenna (6DMA) has been proposed \cite{6DMA_1}. By adjusting arrays' 3D positions/rotations to align their radiation patterns with dominant propagation directions, 6DMA enables statistical channel improvements with practically low movement overhead \cite{6DMA_2}. However, position and rotation variables are tightly coupled in both the optimization objective and physical movement constraints in 6DMA design \cite{6DMA_1}, \cite{6DMA_2}. To alleviate this complexity, a hierarchically tunable 6DMA (HT-6DMA) architecture was introduced in \cite{HT-6DMA} to decouple the optimization of array positions and rotations, achieving high-quality solutions yet with significantly reduced computational complexity.

Despite the above progress in 6DMA, existing architectures rely solely on mechanical means to adjust the position/orientation of antenna arrays.
Such mechanical reconfiguration typically exhibits a latency depending on the antenna/array's movement speed \cite{time} and incurs non-negligible energy consumption during movement, which may also introduce potential interference among arrays during their movement \cite{6DMA_1}-\cite{HT-6DMA}. In contrast, pattern reconfigurable antennas (PRAs) can switch the radiation gain pattern over a discrete mode set \cite{Ref3DTilt} much faster on the order of nanoseconds \cite{time}, also incurring much lower power consumption compared with radio frequency (RF) transmission and processing \cite{PRA-energy}. However, PRA-based beam reconfiguration is practically limited by a finite steering range, e.g., it is unable to cover the area on the backside of the array and fails to cover end-fire directions \cite{RefEndfire}.
To effectively combine the complementary advantages of mechanically-driven 6DMAs and electronically-driven PRAs, we propose in this letter a new hybrid mechanically and electronically tunable 6DMA architecture, termed as HMET-6DMA. Specifically, the base station (BS) equipped with HMET-6DMA consists of multiple antenna arrays which are mechanically movable along a circular rail, while each array is further equipped with PRAs that enable fast adjustment of the radiation gain pattern.
This hybrid architecture retains the spatial adaptability of 6DMA while enhancing its beam adaptation speed via electronic radiation gain pattern reconfiguration.
To maximize the average sum-rate of users, we formulate a two-timescale optimization problem that jointly designs array positions and PRA pattern selections based on the user spatial distribution (assumed to be {\it a priori} known). Numerical results demonstrate that the proposed HMET-6DMA BS architecture outperforms various benchmark schemes in terms of average sum-rate.

\vspace{-0.45 cm}
\section{System Model and Problem Formulation}
\begin{figure}
\centering
\includegraphics[width = 3.3 in, height= 2.4 in]{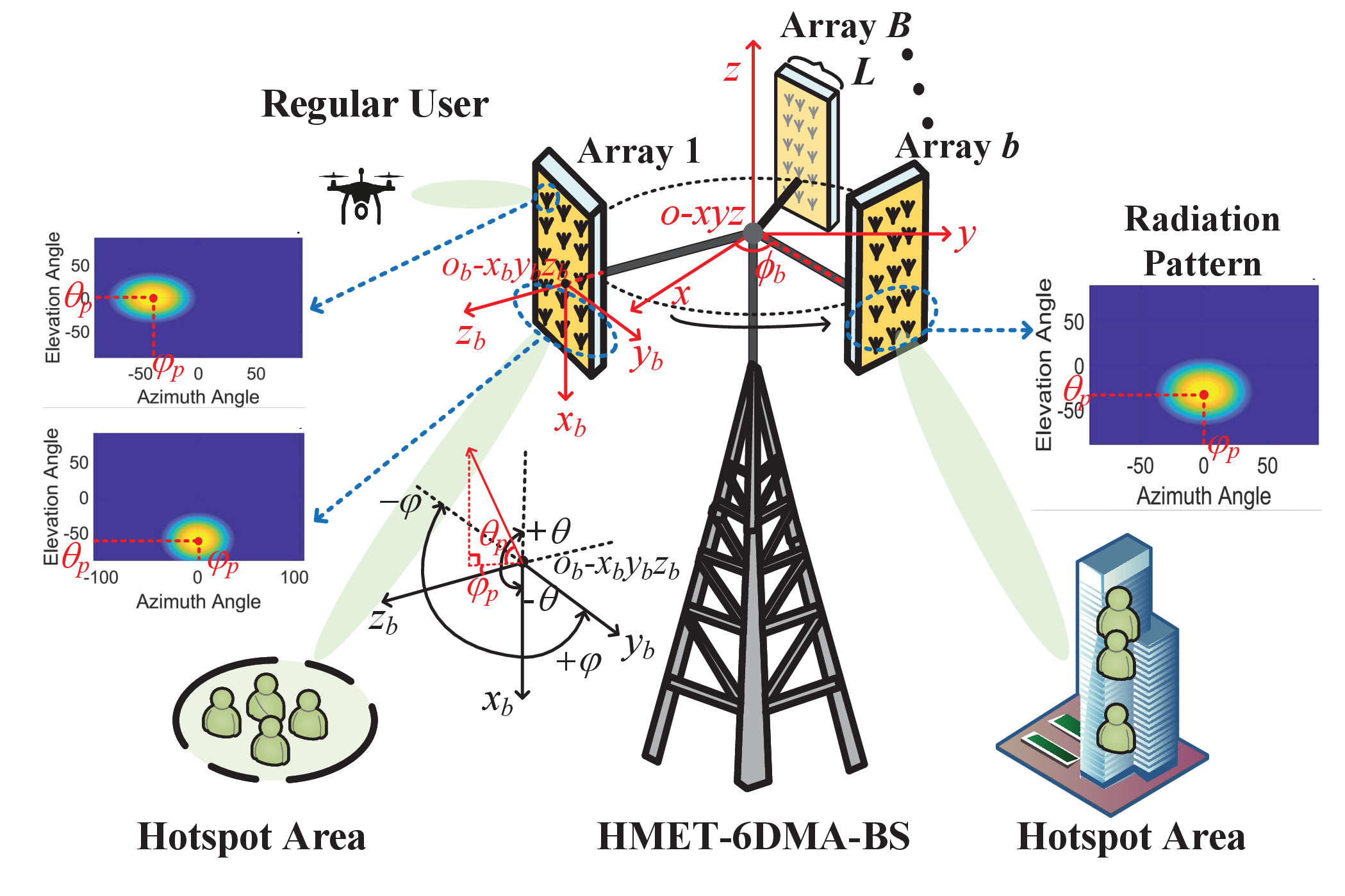}
\vspace{-0.2cm}
\caption{Communication system with the HMET-6DMA BS.}
\vspace{-0.6cm}
\end{figure}

\vspace{-0.25 cm}
\subsection{HMET-6DMA BS Architecture}
\vspace{-0.15 cm}
As shown in Fig. 1, the proposed HMET-6DMA BS consists of $B$ uniform planar arrays (UPAs) indexed by $\mathcal{B}=\{1,\dots,B\}$. Each array has a physical width of $L$ and is mounted on a horizontal circular rail of radius $R$ centered at the BS's center. We assume that each array can be independently moved along the rail by a mechanical means (e.g., stepper motor). The position of the $b$-th array is specified by the azimuth angle $\phi_b \in (-\pi,\pi]$ measured in the global spherical coordinate system (SCS), whose origin coincides with the BS's center, as shown in Fig. 1. The unit outward normal vector of the $b$-th array is denoted by $\mathbf{n}_{b} (\phi_b) = [\cos \phi_b, \sin \phi_b, 0]^T$ in the global Cartesian coordinate system (CCS), which is perpendicular to the horizontal circular rail and shares the same origin as the global SCS.
Under this setup, any two arrays must maintain a minimum angular separation $\beta = 2 \arctan\left( {L}/({2R}) \right)$ to avoid overlap, i.e.,
\begin{equation}
\min\left( |\phi_i - \phi_j|,\ 2\pi - |\phi_i - \phi_j| \right) \geq \beta, \forall i,j \in \mathcal{B}, i \neq j. \label{eq:pos_constraint}
\end{equation}

Moreover, each UPA is composed of $N$ PRAs indexed by $\mathcal{N}=\{1,\dots,N\}$, each of which is electronically tunable, i.e., it is able to switch its radiation gain pattern among a discrete mode set $\mathcal{P} = \{1,2, \cdots, P\}$ to steer its main beam towards different spatial directions. Here we adopt a PRA model that supports 3D beam switching \cite{Ref3DTilt}. For each radiation mode $p$, the main lobe is steered toward the angular pair
$(\theta_p,\varphi_p)$, where $\theta_p$ and $\varphi_p$ denote the elevation and azimuth angles with respect to (w.r.t.) the array's center in its local SCS, respectively, as shown in Fig. 1.
The allowable steering ranges are $\theta_p\in[-\theta_{\max}, \theta_{\max}]$ with $\theta_{\max} \in [\frac{\pi}{6},\frac{\pi}{3}]$ \cite{Yan2019}, and $\varphi_p\in[-\frac{\pi}{2},\frac{\pi}{2}]$. Considering the practical finite angular resolution, the above intervals are uniformly sampled with steps $\Delta\theta$ and $\Delta\varphi$, yielding $\Theta=\{-\theta_{\max},-\theta_{\max}+\Delta\theta, \dots, \theta_{\max} \}, P_v=|\Theta|$ and $
\Pi=\{-\frac{\pi}{2}, -\frac{\pi}{2} +\Delta\varphi,\dots, \frac{\pi}{2} \},P_h=|\Pi|$, with $|\cdot|$ denoting set cardinality and $P = P_h P_v$ the total number of radiation modes. For a mode index $p\in\mathcal{P}$, we have $\theta_p = -\theta_{\max} + \left\lfloor (p-1) / P_h \right\rfloor  \Delta\theta$ and $\varphi_p = -\frac{\pi}{2} + \left[(p-1) \bmod P_h\right]  \Delta\varphi$, where $\bmod$ is the modulo operator and $\lfloor\cdot\rfloor$ denotes the floor function. The radiation pattern of mode $p$ is defined as \cite{3gpp}
\begin{align}
 A_p(\theta, \varphi) = \!G_{\max} \!+\! \Delta G_p \! - \! \min \left( \!-[A_{h,p}(\varphi)+  A_{v,p}(\theta)], G_s \right),
\end{align}
where $G_{\max}$ is the maximum directional gain defined under the default mode with $\theta_p, \varphi_p = 0$ and $\Delta G_p=0$. The functions $A_{h,p}(\cdot)$ and $A_{v,p}(\cdot)$ characterize the azimuth and elevation attenuation of mode $p$, respectively, which are given by
\begin{align}
& A_{h,p}(\varphi)= -\min \Big\{12 \big[ ({\varphi-\varphi_p})/{\varphi_{\text{3dB}}}\big]^2, G_s \Big\}, \\
& A_{v,p}(\theta)= -\min  \Big \{12 \big[ ({\theta-\theta_p})/{\theta_{\text{3dB}}}\big]^2, G_v \Big\}, \nonumber
\end{align}
where $\varphi_{\text{3dB}}$ and $\theta_{\text{3dB}}$ denote the half-power beamwidths in the horizontal and vertical planes, respectively. $G_s$ and $G_v$ denote the front-back ratio and sidelobe level limit, respectively.
Besides, $\Delta G_p=10\log_{10}\!\left(\frac{P_{\mathrm{def}}}{P_{\mathrm{rad}}^{\mathrm{temp}}(p)}\right)$ is a correction term, where $P_{\mathrm{def}}$ is the radiated power of the default mode and $P_{\mathrm{rad}}^{\mathrm{temp}}(p)$ is computed from $A_p(\theta, \varphi)$ with fixed $G_{\max}$. Therefore, the radiated power of mode $p$ is $P_{\mathrm{rad}}(p)=\int_{-\pi}^{\pi}\!\!\int_{-\frac{\pi}{2}}^{\frac{\pi}{2}}
10^{A_p(\theta,\varphi)/10}\, \cos\theta \,\mathrm d\theta\,\mathrm d\varphi=10^{\frac{\Delta G_p}{10}}P_{\mathrm{rad}}^{\mathrm{temp}}(p)=P_{\mathrm{def}}$, $\forall p \in \mathcal{P}$, ensuring identical radiated power across all modes.

\vspace{-0.4cm}
\subsection{Channel Model}
We consider the uplink transmission where $K$ distributed users, each equipped with a single omni-directional antenna, communicate with the HMET-6DMA BS simultaneously. Let $\omega_k \in [-\frac{\pi}{2},\frac{\pi}{2}]$ and $\psi_k \in [-\pi, \pi]$ denote the elevation angle and azimuth angle from the $k$-th user to the center of the BS, respectively. For the $b$-th antenna array, the relationship between the global position of its $n$-th element $\bm r_{b,n}^{\,g}$ and its corresponding position in the local CCS $\bm r_{b,n}^{\,l}$ is given by \cite{HT-6DMA}
\begin{equation}
\bm r_{b,n}^{\,g}=R\, \mathbf{n}_b(\phi_b)+\bm R(\phi_b)\,\bm r_{b,n}^{\,l}, \,\, \forall n \in \mathcal{N}, \forall b \in \mathcal{B},
\end{equation}
where
$\bm R(\phi_b)=
[0, -\sin\phi_b, \cos\phi_b;
0,  \cos\phi_b, \sin\phi_b;
-1,   0, 0]$
maps the coordinate in the array's local CCS to that in the global CCS.
Then the steering vector of the $b$-th array corresponding to the $k$-th user is given by
\begin{equation}
\mathbf{a}_k(\phi_b;\mathbf{f}_k)=
\bigl[e^{-j\frac{2\pi}{\lambda}\mathbf{f}_k^{T}\bm r_{b,1}^{\,g}},\,
      \dots,\,
      e^{-j\frac{2\pi}{\lambda}\mathbf{f}_k^{T}\bm r_{b,N}^{\,g}}\bigr]^{T},
\, b\in\mathcal{B},
\end{equation}
where $\mathbf{f}_k=[\cos\omega_k\cos\psi_k,\ \cos\omega_k\sin\psi_k,\ \sin\omega_k]^{T}$ is the unit pointing vector from the origin to the $k$-th user and $\lambda$ denotes the wavelength of the carrier.

Let the pattern selection matrix of the $b$-th antenna array be denoted by $\mathbf{Z}_b \in \mathbb{Z}^{P \times N}$, with each entry taking the value 0 or 1. The entry at the $p$-th row and $n$-th column, denoted by $z_{p,n}^{(b)}$, indicates whether the $n$-th antenna in the $b$-th array selects the $p$-th pattern (if it equals 1) or not (if it equals to 0). To ensure that each antenna selects exactly one pattern, each column of $\mathbf{Z}_b$ must only contain one entry with 1 and zeros elsewhere, i.e., $\mathbf{1}_P^T \mathbf{Z}_b = \mathbf{1}_N$. Then the antenna gain from the $k$-th user to the $n$-th antenna on the $b$-th array in the linear scale is given by
$g_{b,n,k} = {\sum}_{p=1}^{P} \,\,\, z_{p,n}^{(b)} \cdot 10^{A_p(\tilde{\theta}_k^b, \tilde{\varphi}_k^b)/10}$,
where $\tilde{\theta}_k^b =-\arcsin(\widetilde{x}_{bk})$ and $\tilde{\varphi}_k^b = \arctan2 (\widetilde{y}_{bk},\widetilde{z}_{bk})$ represent the elevation and azimuth angles from the $k$-th user to the center of the $b$-th array in its local SCS. Here, the vector $[\widetilde{x}_{bk},\widetilde{y}_{bk},\widetilde{z}_{bk}]^T = \bm R(\phi_b)^{-1} \mathbf{f}_{k}$ denotes the unit-length pointing vector in the $b$-th array's local CCS. For convenience, a sparse matrix $\mathbf{M}_{b,k} \in \mathbb{R}^{N \times PN}$ is introduced such that $[\mathbf{M}_{b,k}]_{n,\,(n-1)P + p} = \sqrt{10^{A_p(\tilde{\theta}_k^b, \tilde{\varphi}_k^b)/10}}, \forall n \in \mathcal{N}, \forall p \in \mathcal{P}$, where all the other entries are zeros. For the ease of exposition, we assume the line-of-sight (LoS) far-field channel model between each user and the BS\footnote{The channel model can be easily extended to the general multi-path channel model with small-scale channel fading.}. Then the uplink channel from the $k$-th user to the $b$-th antenna array is expressed as
\begin{equation}
\mathbf{h}_{b,k}(\phi_b, \mathbf{z}_{b}; \mathbf{f}_{k}) \! = \! \sqrt{v_{k}}  \operatorname{diag}[\mathbf{a}_k(\phi_b; \mathbf{f}_{k})]  \mathbf{M}_{b,k}(\phi_b; \mathbf{f}_{k}) \mathbf{z}_b,
\end{equation}
where $v_{k} = {\epsilon}_{0} d_{k}^{-\varsigma}$ represents the channel power gain of the $k$-th user, with ${\epsilon}_{0}$ being a constant, $d_{k}$ the distance between the $k$-th user and the BS's center, $\varsigma$ the path-loss exponent, and $\mathbf{z}_b = \operatorname{vec}(\mathbf{Z}_b)$ denoting the column-wise vectorization of the matrix $\mathbf{Z}_{b}$. Let $\bm{\phi}= [\phi_1, \cdots,\phi_B]^T$ denote the stacked variables. Then the overall channel vector from the $k$-th user to all $B$ antenna arrays is obtained as $\mathbf{h}_k (\bm{\phi},\mathbf{z}; \mathbf{f}_{k}) = [\mathbf{h}^T_{1,k}(\phi_1, \mathbf{z}_{1}; \mathbf{f}_{k}), \cdots, \mathbf{h}^T_{B,k}(\phi_B, \mathbf{z}_{B}; \mathbf{f}_{k})]^T$, where $\mathbf{z} \! = \! [\mathbf{z}_1^T, \dots, \mathbf{z}_B^T]^T$ is the overall pattern selection vector.

Accordingly, the received signal at the BS is expressed as
$\mathbf{y} = \mathbf{H}(\bm{\phi}, \mathbf{z}; \mathbf{F})\mathbf{x} + \mathbf{n}$, where $\mathbf{x} = [x_1, x_2, \dots,x_K]^T \in \mathbb{C}^{K \times 1}$ is the transmitted signal vector with $x_k$ denoting the signal from the $k$-th user with average power $p_k, k =1,\cdots,K$. The multi-user channel matrix $\mathbf{H}(\bm{\phi}, \mathbf{z}; \mathbf{F}) = [\mathbf{h}_1(\bm{\phi}, \mathbf{z}; \mathbf{f}_{1}), \mathbf{h}_2(\bm{\phi}, \mathbf{z}; \mathbf{f}_{2}), \dots, \mathbf{h}_K(\bm{\phi}, \mathbf{z}; \mathbf{f}_{k})] \in \mathbb{C}^{NB \times K}$ constitutes the channels from all the $K$ users to all the $NB$ array antennas at the HMET-6DMA BS, with $\mathbf{F} \triangleq [\mathbf{f}^T_{1},\cdots,\mathbf{f}^T_{K}]^T$. The vector $\mathbf{n} \sim \mathcal{CN}(\mathbf{0}_{NB}, \sigma^2 \mathbf{I}_{NB})$ represents the complex additive white Gaussian noise (AWGN) at the BS with zero mean and average power $\sigma^2$ at each receive antenna.

\vspace{-0.3cm}
\subsection{Problem Formulation}\label{subsec:prob_form}
\vspace{-0.15cm}
Assuming perfect channel state information (CSI) at the BS and adopting Gaussian signalling with joint multi-user decoding, the achievable average sum-rate of all users is \cite{6DMA_1}
\begin{align}
& C(\bm{\phi}, \mathbf{z}) = \label{eq:obj}\\
&\mathbb{E}_{\mathbf{H}}\bigg[ \log_{2}\det\big(\mathbf{I}_{NB} + \frac{ p_k }{\sigma^{2}} {\sum}_{k=1}^{K} \mathbf{h}_k(\bm{\phi},\mathbf{z};\mathbf{f}_{k})\, \mathbf{h}_k^{H}(\bm{\phi},\mathbf{z};\mathbf{f}_{k})\big)\bigg]. \nonumber
\end{align}
The expectation is taken over the random channel matrix $\mathbf{H}$, which varies with the user locations under our considered LoS channel model. Since it is difficult to express (\ref{eq:obj}) further analytically, we approximate the expectation via Monte Carlo sampling. Specifically, we generate $S$ independent samples of $\mathbf{H}$ according to the given user location/channel distributions (to be specified in Section IV), each containing $K_s$ randomly generated users and their positions in terms of the unit pointing vector from the origin to the $k$-th user in the $s$-th channel sample $\{\mathbf{f}_{k,s}\}_{k=1}^{K_{s}}$ and the corresponding user-BS distances in the $s$-th channel sample $\{d_{k,s}\}_{k=1}^{K_{s}}$, for $s\in \mathcal{S} \triangleq \{1,\dots,S\}$.
Then the corresponding channel vector of the $k$-th user in the $s$-th sample is represented as $\mathbf{h}_{k,s}(\bm{\phi},\mathbf{z}_{s})\triangleq \mathbf{h}_{k}(\bm{\phi},\mathbf{z};\mathbf{f}_{k,s}, d_{k,s})$. Thus, the average sum-rate in (\ref{eq:obj}) is approximated by
\begin{align}
&\hat{C}(\bm{\phi}, \{\mathbf{z}_{s}\}_{s=1}^{S}) = \label{eq:obj_MC} \\
&\frac{1}{S} {\sum}_{s=1}^{S} \log_2 \det \bigg(\mathbf{I}_{NB}  +  \frac{p_k}{\sigma^2} {\sum}_{k=1}^{K_s}\mathbf{h}_{k,s}(\bm{\phi}, \mathbf{z}_{s}) \mathbf{h}^H_{k,s}(\bm{\phi}, \mathbf{z}_{s}) \bigg),\nonumber
\end{align}
where $\mathbf{z}_{s}$ denotes the antenna radiation pattern selection vector in the $s$-th channel sample on the short timescale, while $\bm{\phi}$ denotes the array position vector across all the $S$ channel samples on the long timescale.

Our aim is to maximize the approximated average sum-rate in (\ref{eq:obj_MC}) via a two-timescale optimization over $\bm{\phi}$ and $\{\mathbf{z}_{s}\}_{s=1}^{S}$. The problem is thus formulated as
\begin{subequations}
\begin{align}
&\!\!(\text{P}1): \max_{ \bm{\phi},\{\mathbf{z}_{s}\}_{s=1}^{S}} \, \hat{C}(\bm{\phi},\{\mathbf{z}_{s}\}_{s=1}^{S}) \\
&\!\!\text{s.t. } ( \mathbf{1}_P^T \otimes \mathbf{I}_N ) \cdot \mathbf{z}_{b,s} = \mathbf{1}_N, \,\, \forall b \in \mathcal{B}, \,\, \forall s \in \mathcal{S}, \label{cons1}\\
&\!\! \mathbf{z}_{b,s}(u) \in \{0,1\},\,\, \forall u = 1, 2, \dots, PN, \,\, \forall s \in \mathcal{S}, \label{cons2} \\
&\!\!\!\min\left( |\phi_i - \phi_j|,\ 2\pi - |\phi_i - \phi_j| \right) \geq \beta, \forall i,j  \in \mathcal{B}, j\neq i, \label{cons4}
\end{align}
\end{subequations}
where $\otimes$ stands for the kronecker product. In (P1), $\mathbf{z}_{b,s}$ denotes the pattern selection vector for the $b$-th array in the $s$-th channel sample. Constraint (\ref{cons1}) enforces that each antenna chooses exactly one radiation pattern, (\ref{cons2}) imposes the binary decision constraint, and (\ref{cons4}) guarantees the minimum angular separation $\beta$ between any two arrays to prevent overlap.

\vspace{-0.2cm}
\section{Proposed Algorithm}
\vspace{-0.1cm}
In this section, the joint optimization framework is presented. Due to the different time scales and system constraints associated with the design variables, we adopt a two-timescale block coordinate ascent (BCA) optimization strategy. The long-timescale optimization for $\phi$ is implemented based on the user distribution statistics, while the antenna radiation pattern selection is designed according to instantaneous user locations on each channel sample $s$.

\vspace{-0.35cm}
\subsection{Long-timescale Optimization for Array Position $\bm{\phi}$}
\vspace{-0.15cm}
We solve the subproblem w.r.t. $\bm{\phi}$ via alternating optimization in the long timescale. Specifically, we optimize each $\phi_b$ with given $\{\phi_j\}_{j \in \mathcal{B} \setminus b}$ in each iteration. The corresponding optimization problem in each iteration is expressed as
\begin{subequations}
\begin{align}
&\!\!\!\!\!\!\!\!\!\! (\text{P}2):\max_{ {\phi}_{b}} \, \hat{C}({\phi}_{b}) \\
\text{s.t. }
& \min\left( |\phi_b - \phi_j|,\ 2\pi - |\phi_b - \phi_j| \right) \geq \beta, \forall j  \in \mathcal{B} \setminus b.
\end{align}
\end{subequations}
Therefore, the feasible region for the position of $b$-th antenna array $\phi_{b}$ is thus defined as
${\mathcal{F}}_{b} =  \big\{ \phi_b \in (-\pi, \pi]\, \big |
\left| \left( (\phi_b - \phi_j + \pi) \bmod 2\pi \right) - \pi \right| \geq \beta,\; \forall j \in \mathcal{B} \setminus b \big\}$.
Since the constraint is fully captured by the feasible set $\mathcal{F}_b$, (P2) reduces to an optimization problem over $\mathcal{F}_b$ without additional constraints. To solve this problem, we adopt the projected gradient ascent method. At each iteration $t$, the update of $\phi_b^{(t)}$ is given by
$\phi_b^{(t)} = \text{proj}_{\mathcal{F}_b}\left( \phi_b^{(t-1)} + \eta \triangledown \hat{C}(\phi^{(t-1)}_b)  \right)$,
where $\triangledown \hat{C}(\phi^{(t-1)}_b)$ denotes the gradient with given $\phi^{(t-1)}_b$ at iteration $t-1$, and $\eta$ is the step size which is set such that $\phi_b^{(t)}$ satisfies the Armijo condition \cite{convex}. In addition, $\text{proj}_{\mathcal{F}_b}(\cdot)$ denotes the projection operator that maps the tentative value back to the feasible region $\mathcal{F}_b$.

The inner iteration terminates if $|\hat{C}(\phi_b^{(t)}) - \hat{C}(\phi_b^{(t-1)})| < \epsilon_{\text{th}}$ is satisfied, where $\epsilon_{\text{th}}$ is a small positive threshold, or when the number of inner iterations reaches a predefined maximum $T_{\text{in},1}$. In the outer loop, the alternating updates over $\{\phi_j\}_{j \in \mathcal{B}}$ are repeated until the objective value converges or the outer iteration count exceeds a predefined limit $T_{\text{out},1}$.

\vspace{-0.3cm}
\subsection{Short-timescale Optimization for Pattern Selection $\{\mathbf{z}_{s}\}$}
\vspace{-0.1cm}
Next, we focus on the short-timescale pattern selection $\{\mathbf{z}_{s}\}_{s \in \mathcal{S}}$ based on instantaneous channel samples with given $\bm{\phi}$. Since different channel samples are independent, the optimization can be decoupled over different channel realizations.

For each channel sample indexed by $s \in \mathcal{S}$, we apply an iterative greedy-based search for a near-optimal pattern selection vector. In each outer iteration ($t = 1, 2, \cdots, T_{\text{out},2}$), the arrays are sequentially updated in the order of $b = 1, \cdots, B$. When array $b$ is selected for update, its pattern-selection vector in the $s$-th sample $\mathbf{z}_{b,s}$ is updated while keeping those of the other arrays fixed. We perform only a single-pass update for each array. Specifically, for the $n$-th antenna in array $b$ in the $s$-th sample, the corresponding pattern selection vector $\mathbf{z}_{b,n,s} \in \mathbb{R}^{P \times 1}$ is updated sequentially over antennas, and the
corresponding optimal pattern index $p^{\star}_{b,n,s} \in \mathcal{P}$ is chosen to maximize the objective function $\hat{C}(\cdot)$, i.e.,
\begin{equation}
p^{\star}_{b,n,s} = \arg \max \,\,\, \hat{C}(\mathbf{z}_{b,n,s} = \mathbf{e}_p), \,\,\,{\forall p \in \mathcal{P}},
\end{equation}
where $\mathbf{e}_p$ denotes the $p$-th standard basis vector in $\mathbb{R}^P$.

After one full sweep over all $B$ arrays, the iteration counter $t$ is increased by one and a complete pattern selection vector $\mathbf{z}^{(t)}_{s}$ is obtained. The procedure with $S$ channel samples carried out independently stops if the objective value converges, or when the maximum number of outer iterations $T_{\text{out},2}$ is reached.

The algorithm alternates between the array position block and the pattern selection block and stops when the relative objective improvement is below $\epsilon_{\text{th}}$ or the maximum outer cycle $T_{\text{cycle}}$ is reached. The overall algorithm to solve the problem (P1) is shown in Algorithm 1.
The computational complexity of the proposed algorithm is $\mathcal{O}\!\left( T_{\text{out},1}T_{\text{in},1}  B S K_s^{3} + T_{\text{out},2} S B N P K_s^{3} \right)$ in the worst-case scenario. In practice, the iterations may terminate earlier when the relative objective improvement falls below $\epsilon_{\text{th}}$.
Since the gradient/greedy-based optimization approach always leads to non-decreasing objective values and the optimal value is upper bounded, the convergence is ensured.

\begin{small}
\begin{algorithm}[!t]
\caption{BCA-based Joint Position and Pattern Optimization}
\label{alg:BCA}
\begin{algorithmic}[1]
\STATE \textbf{Input:} $B,N,S,P,T_{\text{cycle}},T_{\text{in},1},T_{\text{out},2},
       \eta_{\mathrm{init}},\delta,\epsilon_{\text{th}}$
\STATE Initialize feasible $\bm{\phi}^{(0)}$ and $\{\mathbf z_{b,s}^{(0)}\}$; set cycle index $r=0$, $m_1 = 1$, $m_2 = 0$
\REPEAT
        \WHILE{$m_1 < T_{\text{out},1}$}
    \FOR{$b=1$ to $B$}
        \STATE Fix $\{\phi_j\}_{j\neq b}$, set $t=1$ and $\phi_b^{(0)}\gets\phi_b$
        \WHILE{$t < T_{\mathrm{in},1}$}
            \STATE Compute gradient $\nabla_{\phi_b}\hat C(\phi_b^{(t-1)})$ and set stepsize $\eta^{(t-1)} \gets \eta_{\mathrm{init}}$
            \STATE Tentative update $\phi_b^{(t)} = \text{proj}_{\mathcal{F}_b}\left( \phi_b^{(t-1)} + \eta \triangledown \hat{C}(\phi^{(t-1)}_b)  \right)$
            \WHILE{$\hat C(\phi_b^{(t)})-\hat C(\phi_b^{(t-1)}) < \ell\,\eta^{(t-1)}\,\triangledown \hat{C}(\phi^{(t-1)}_b)(\phi_{b}^{(t)}-\phi_b^{(t-1)}) $}
                \STATE $\eta^{(t)} \gets \delta\,\eta^{(t-1)} $; $\phi_b^{(t)} = \text{proj}_{\mathcal{F}_b}\left( \phi_b^{(t-1)} + \eta \triangledown \hat{C}(\phi^{(t-1)}_b)  \right)$
            \ENDWHILE
            \IF{$|\hat C(\phi_b^{t+1})-\hat C(\phi_b^{t})| \le \epsilon_{\mathrm{th}}$}
                \STATE   $t \gets t+1$, \textbf{break}
            \ENDIF
            \STATE $t \gets t+1$
        \ENDWHILE
    \ENDFOR
    \STATE $m_1 \gets m_1 +1$
 \ENDWHILE
    \FORALL{channel samples $s\in\mathcal S$ \textbf{in parallel}}
        \REPEAT
            \FOR{$b=1$ to $B$}
                \FOR{$n=1$ to $N$}
                    \STATE
                    $p^{\star}_{b,n,s} \gets
                    \arg\max_{p\in\{1,\dots,P\}}
                    \hat C(\mathbf z_{b,n,s}=\mathbf e_p)$;
                \ENDFOR
            \ENDFOR
            \STATE $m_2 \gets m_2+1$.
        \UNTIL{$m_2 \ge T_{\mathrm{out},2}$ or relative improvement $\le \epsilon_{\mathrm{th}}$}
    \ENDFOR
    \STATE $r \gets r+1$.
\UNTIL{$r \ge T_{\mathrm{cycle}}$ or relative improvement of $\hat C(\bm{\phi}^{(r)},\{\mathbf z_{b,s}^{(r)}\})$ $\le \epsilon_{\mathrm{th}}$}
\STATE \textbf{Output:} $\bm{\phi}^{\star}=\bm{\phi}^{(r)}$, $\{\mathbf z_{b,s}^{\star}\}=\{\mathbf z_{b,s}^{(r)}\}$.
\end{algorithmic}
\end{algorithm}
\end{small}

\begin{figure}
\centering
\includegraphics[width = 2.5 in, height= 1.6 in]{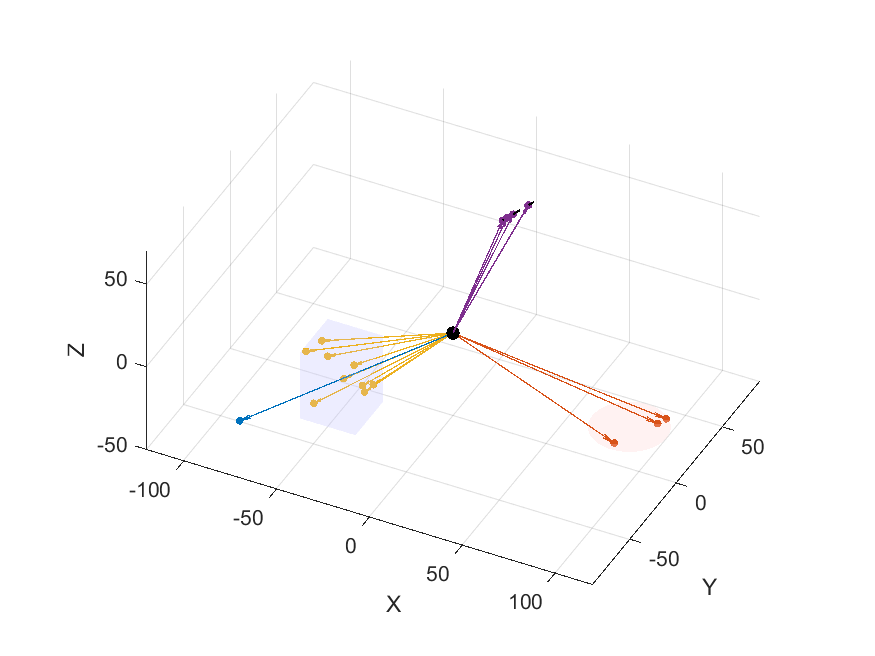}
\vspace{-0.2cm}
\caption{User spatial distribution ($\eta = 0.3$).}
\vspace{-0.4cm}
\label{fig:conv}
\end{figure}

\section{Simulation Results}
In this section, numerical results are provided to evaluate the performance of the proposed HMET-6DMA BS design, as compared to various benchmark schemes. Unless otherwise specified, we consider each BS array as a UPA with half-wavelength antenna spacing, and the radius of the circular rail is set to $R = 1~\mathrm{m}$. We set $\theta_{\max}=\pi/3$, and the PRA steering resolution to $\Delta\theta=\Delta\varphi=\pi/12$. The PRA radiation pattern parameters are set to $G_{\max}=8$~dBi, $\theta_{3\mathrm{dB}}=\varphi_{3\mathrm{dB}}=30^\circ$, $G_s=30$~dB, and $G_v=30$~dB. The carrier frequency is $f_c = 2.4~\mathrm{GHz}$, the transmit power per user is $p_k = p = 30~\mathrm{mW}, \forall k = 1, \cdots,K$, the noise power is $\sigma^2 = -50~\mathrm{dBm}$, and the number of Monte Carlo channel samples is $S = 100$. The minimum angular separation $\beta$ is set to $\pi/24$ to avoid physical overlap. For the proposed algorithms, the convergence threshold is $\epsilon_{\mathrm{th}} = 5 \times 10^{-4}$, the maximum inner and outer iterations for array position optimization are set to $T_{\text{in},1}=50$ and $T_{\text{out},1}=3$, respectively, and the maximum iterations for the design of the pattern selection vector are $T_{\text{out},2}=3$. In addition, the maximum outer cycle is set to $T_{\text{cycle}} = 2$.

\begin{figure}
\centering
\includegraphics[width = 2.1 in, height=1.8 in]{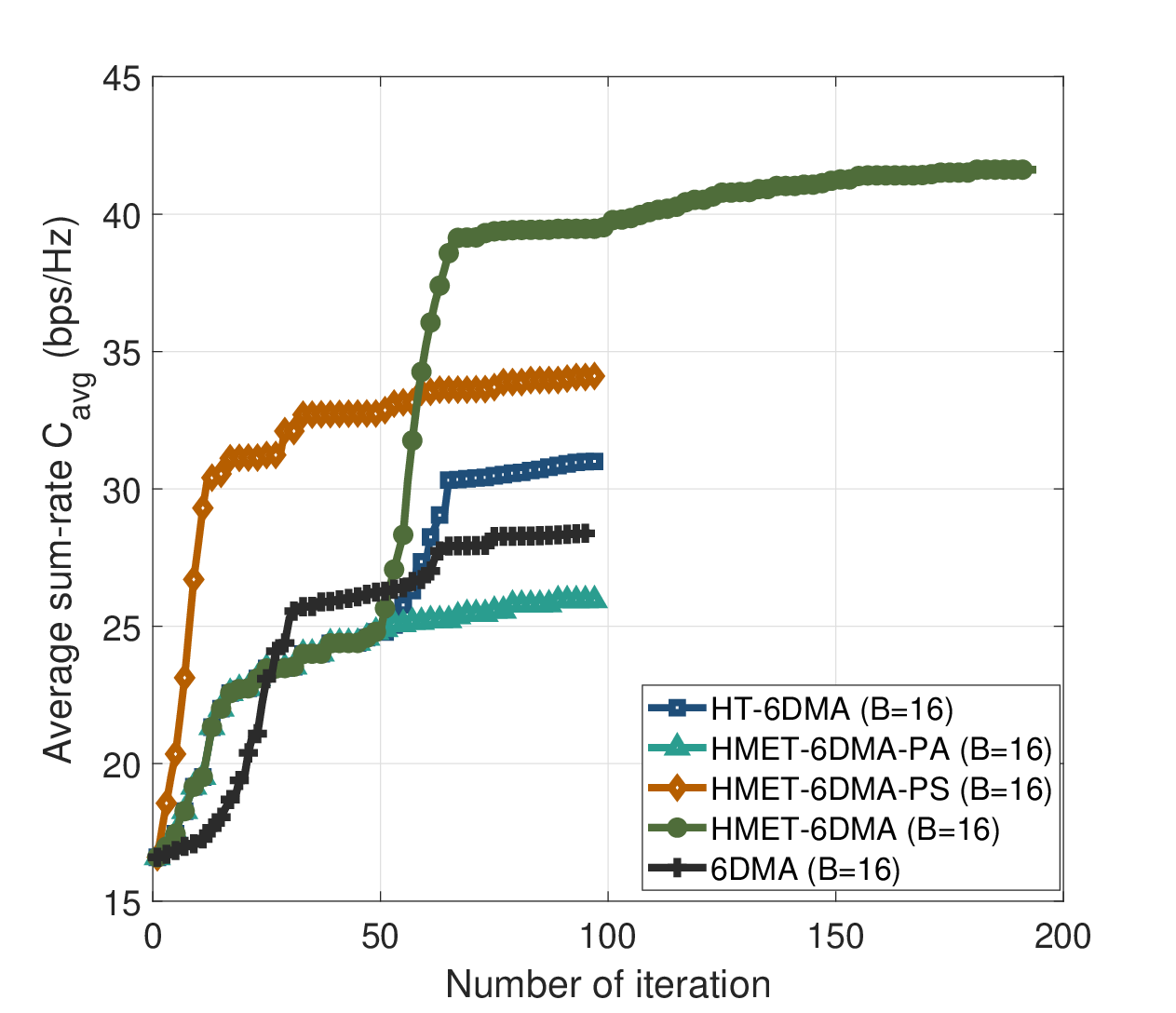}
\vspace{-0.1cm}
\caption{Convergence behavior.}
\vspace{-0.6cm}
\label{fig:conv}
\end{figure}

\subsection{Performance under time-invariant user distributions}
We adopt a non-homogeneous Poisson process (NHPP) to model user locations \cite{HT-6DMA}, as shown in Fig. 2. The coverage region of the BS is a 3D spherical space denoted by $\mathcal{A}$, centered at the BS with radial distance ranging from 50 to 120 m. Within $\mathcal{A}$, three hotspot subregions are defined: the building region $\mathcal{A}_{1}$ is a cuboid centered at (-50, -20, -30) m with size 30 $\times$ 30 $\times$ 40 m, the ground region $\mathcal{A}_{2}$ is a circular area at $z = -50$ centered at (80, 30, -50) m with radius 20 m, and the airway region $\mathcal{A}_{3}$ is a line segment from (10, 30, 40) m to (30, 30, 70) m. Then, the regular-user region $\mathcal{A}_{0}$ denotes the remaining coverage area excluding the above hotspots. For each sample, the number of users located in region $\mathcal{A}_{i}$ is denoted by $K_i$, which follows a Poisson distribution, i.e.,
\begin{equation}
 \Pr[K_i  =  \tilde{K}]  =  {\mu_i^{\tilde{K}} e^{-\mu_i}} / ({\tilde{K}!}), \tilde{K}  =  0,1,\ldots, i \in \{0,1,2,3 \},
\end{equation}
where $\mu_{i}$ denotes the mean. We assume $\mu_{1} = \mu_{2} = \mu_{3}$ and the users are uniformly distributed within each region. The total mean number of users is set to $\mu_{t}=\sum_{i=0}^{3} \mu_i = 24$. The sparsity ratio is defined as the average regular/non-hotspot user number over the total average user number, i.e., $\eta=\mu_0/\mu_{t}$, which quantifies the sparsity level of the user distribution. We consider two settings with different numbers of antenna arrays, i.e., $B=3$ and $B=16$, while keeping the total number of antennas fixed.

For performance comparison, the following benchmarks are considered:
i) \textbf{FPA}: The BS has three fixed arrays equally spaced on the rail, each with $\lfloor \tfrac{NB}{3}\rfloor$ antennas, and each antenna operates with its beam directed straight ahead ($\varphi_p = \theta_p = 0^{\circ}$);
ii)-iii) \textbf{6DMA} \cite{6DMA_1}/\textbf{HT-6DMA} \cite{HT-6DMA}: Both array positions and rotations are mechanically controlled, with $B$, $N$, and $\mathcal{C}$ set to be the same as those for the HMET-6DMA BS;
iv) \textbf{HMET-6DMA BS with position adjustment (PA) only}: Only the array positions are optimized on the long timescale, with fixed antenna beam directed straight ahead;
\textbf{HMET-6DMA BS with pattern selection (PS) only}: The array positions are fixed and equally placed along the circular rail, and the design of PRAs is optimized on the short timescale.

Fig. 3 shows the convergence behaviors of the proposed algorithms for HMET-6DMA, with $(B,N) = (16, 4)$ and $\eta = 0.4$ unless specified. For fair comparison, all 6DMA-based schemes are initialized with antenna arrays equally spaced on the circular rail and with the beams directed straight ahead. For the HMET-6DMA scheme, note that the number of iterations corresponds to the number of updates for the position or pattern selection variable of each array, e.g., the first 48 iterations correspond to the array position updates ($B = 16, T_{\text{out},1} = 3$), followed by the second 48 pattern selection updates ($B = 16, T_{\text{out},2} = 3$), under two outer cycles of the update.
For the HT-6DMA scheme, it runs only 96 iterations (48 position updates and 48 orientation updates) \cite{HT-6DMA}.
For the HMET-6DMA-PS ($B=16$) and HMET-6DMA-PA scheme, the maximum iterations for the design of the pattern selection vector and array position are set as $T_{\text{out},1}= T_{\text{out},2}= 6$.
For the 6DMA scheme, the first $B=16$ iterations correspond to the array position updates, followed by the second $B=16$ iterations for rotation updates \cite{6DMA_1}.
It is observed that the proposed HMET-6DMA converges to a higher sum-rate value than HT-6DMA. This performance improvement is attributed to two key factors: (i) mechanical position adjustment aligns the arrays with user distributions in the horizontal plane, while electronic pattern selection further adapts the beams in the vertical domain and provides coverage for sparsely distributed regular users; and (ii) pattern selection is optimized at each channel sample, thus offering higher degrees of freedom to cater to instantaneous variations in user locations.
Besides, the HT-6DMA scheme achieves better performance than 6DMA owing to the hierarchically tunable ability \cite{HT-6DMA}, with array positions and rotations represented in the global and local SCSs, respectively.
In addition, it is observed that the HMET-6DMA with PA only and that with PS only suffer rate losses as compared to the HMET-6DMA with both PA and PS, indicating that both position optimization and pattern selection play significant roles in improving the sum rate. The HMET-6DMA with PA only improves horizontal alignment by exploiting long-term user distribution statistics, but it lacks the flexibility to respond to instantaneous user location/channel variations and cannot provide vertical alignment. In contrast, the HMET-6DMA with PS only provides fast adaptation, yet its performance is limited by the restricted electronic tuning range of PRAs.

The results in Fig.~\ref{fig:conv} are obtained with all 6DMA-based schemes initialized by antenna arrays equally spaced on the circular rail and with the beams directed straight ahead. Since the performance of HMET-6DMA may depend on the initial configurations, we further evaluate the achievable average sum rate under different initializations, including different initial array positions along the rail and different initial PRA state selections, as shown in Fig. 4 and Fig. 5. Across the tested initializations, the same conclusion holds: HMET-6DMA performance is not sensitive to the initialization and the proposed hybrid architecture consistently maintains a clear advantage over all baseline schemes.

\begin{figure}[!t]
		\centering
		\setlength{\abovecaptionskip}{+2mm}
		\setlength{\belowcaptionskip}{+1mm}
		\subfigure[Initial antenna position setup for all 6DMA-based schemes.]
            {\includegraphics[width=1.5in, height = 1.3in]{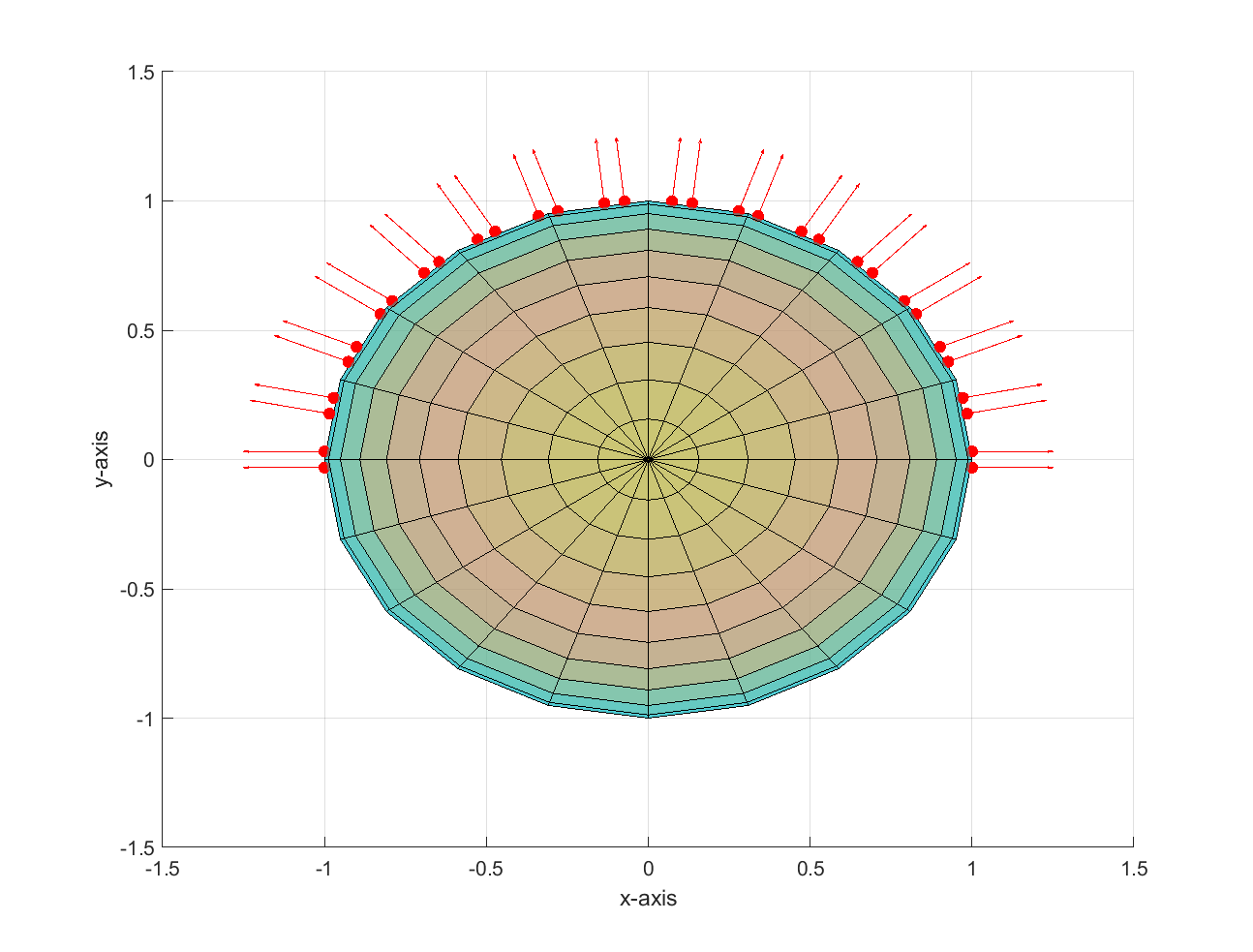}}
        \subfigure[Achievable average sum-rate versus sparsity level of the user distribution $\eta$.]
        {\includegraphics[width=1.5in, height = 1.3in]{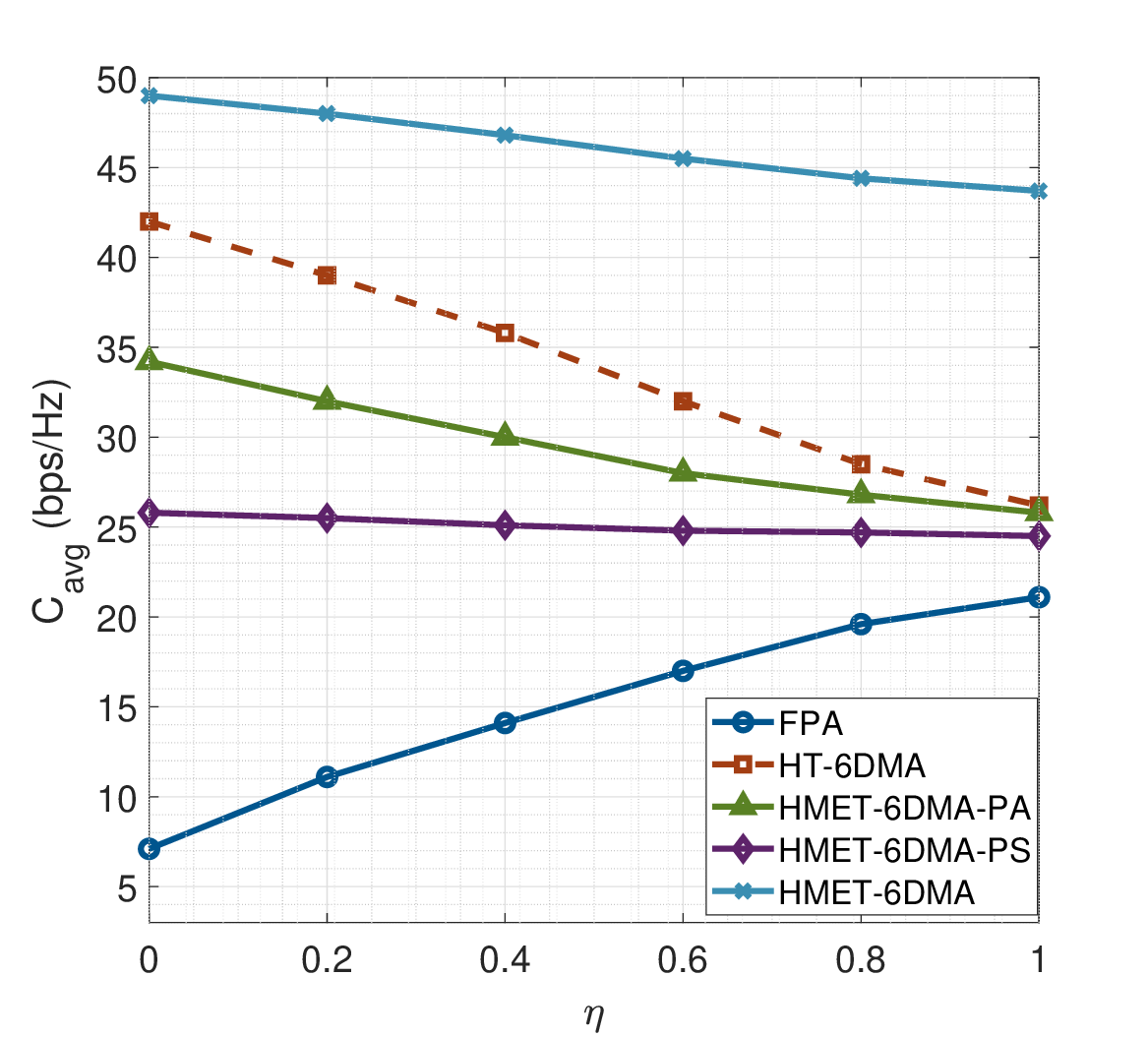}}
		\caption{Optimization results with different initial antenna positions}
\vspace{-0.6cm}
\end{figure}

\begin{figure}[!t]
		\centering
		\setlength{\abovecaptionskip}{+2mm}
		\setlength{\belowcaptionskip}{+1mm}
		\subfigure[Initial antenna radiation pattern setup for all 6DMA-based schemes.]
        {\includegraphics[width=1.5in, height = 1.3in]{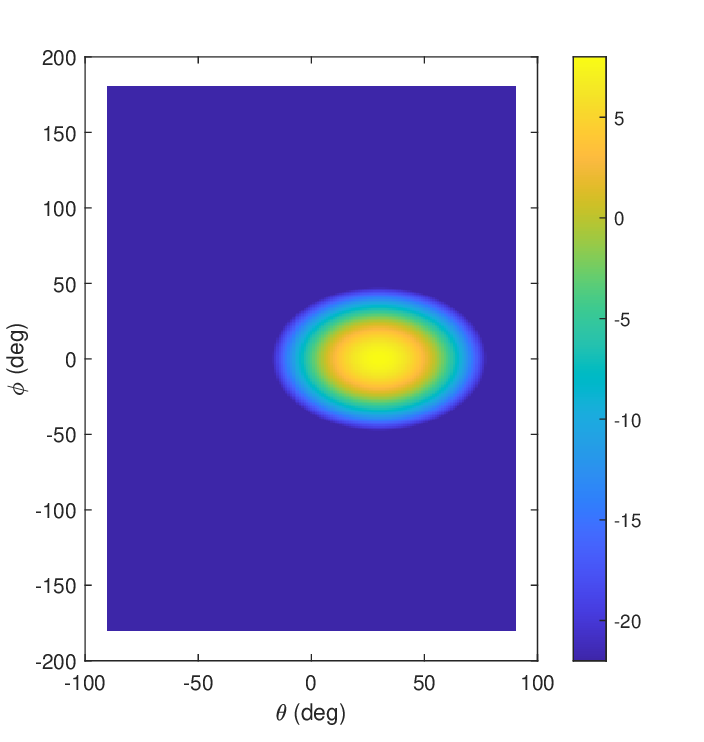}}
        \subfigure[Achievable average sum-rate versus sparsity level of the user distribution $\eta$.]
        {\includegraphics[width=1.5in, height = 1.3in]{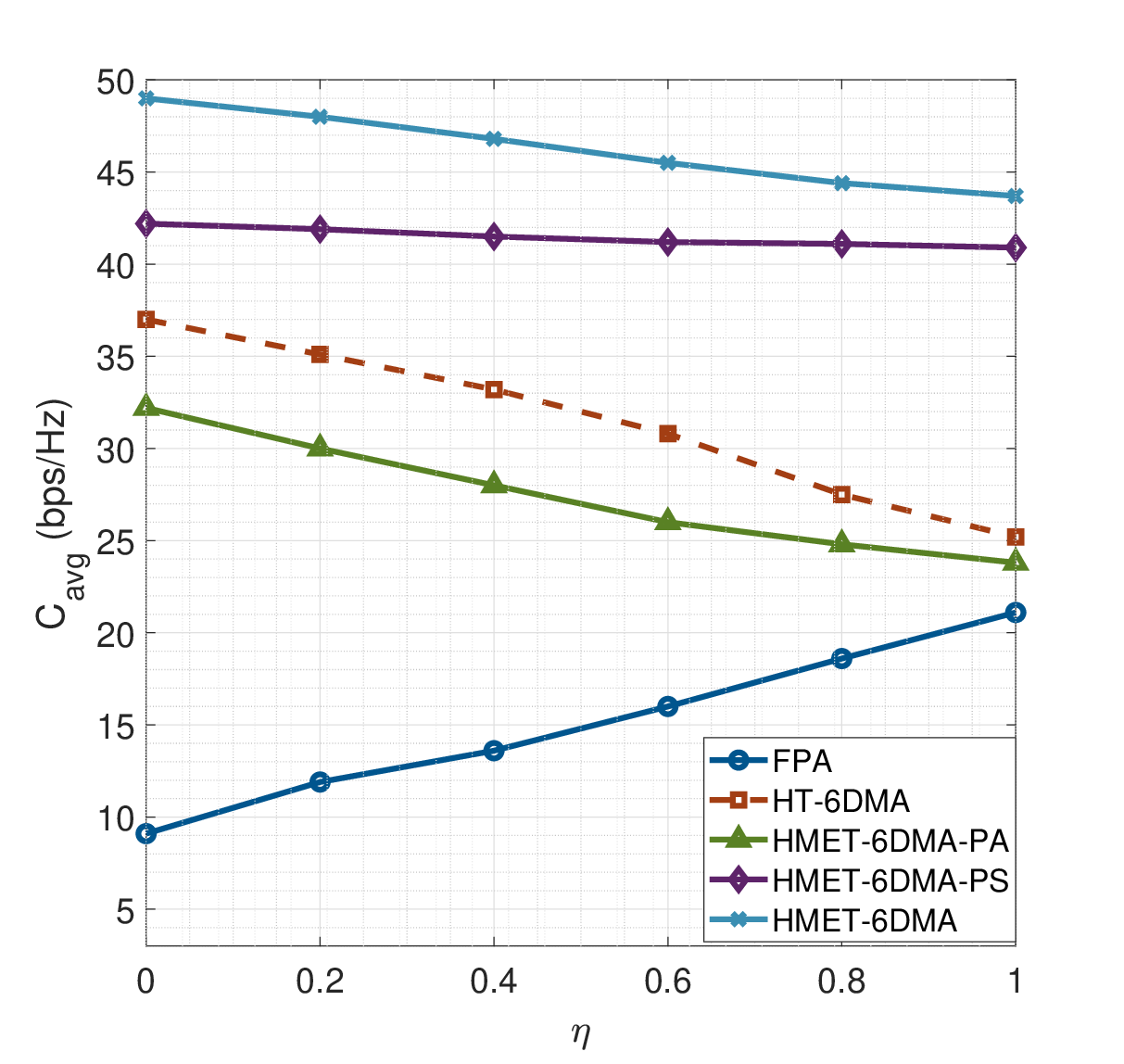}}
		\caption{Optimization results with different initial antenna radiation patterns.}
\vspace{-0.4cm}
\end{figure}

\begin{figure}[!t]
\centering
\includegraphics[width = 1.9 in, height= 1.5 in]{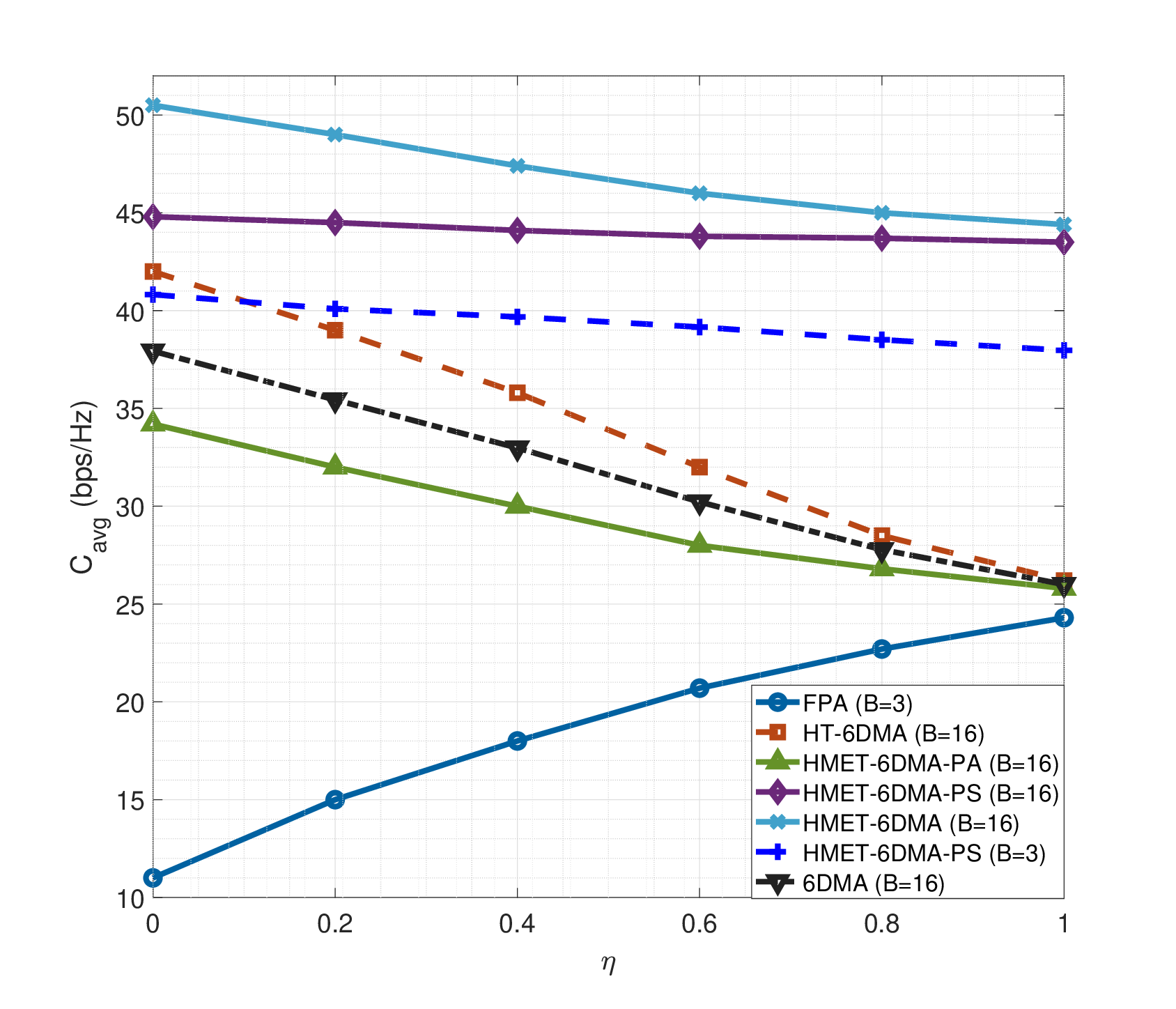}
\vspace{-0.2cm}
\caption{Achievable sum rate versus $\eta$.}
\vspace{-0.5cm}
\label{fig:sparsity}
\end{figure}

Fig. \ref{fig:sparsity} shows the achievable sum rate versus the sparsity ratio of user distribution $\eta$. When $\eta$ increases, the sum rate of the FPA scheme increases while those of all the 6DMA-based schemes decrease. This is due to the fact that when the users are more uniformly distributed, the benefit of 6DMA in flexibly allocating the antenna resources for hotspot areas becomes less significant. Besides, among all the 6DMA-based schemes, those schemes with pattern selection optimization degrade less since the antenna radiation gain pattern is tuned for short-timescale channel adaptation which is more effective than mechanical antenna position adjustment based on the long-timescale user distribution. Finally, with the same number of antenna elements, HMET-6DMA-PS with $B=16$ outperforms that with $B=3$ since it adopts more arrays to provide higher flexibility in electronic pattern selection.

\subsection{Performance under time-varying user distributions}
The results above are obtained under a time-invariant large-scale user geometry, where the hotspot regions are fixed while the user realizations across channel samples are generated in a memoryless manner. To provide a more comprehensive evaluation, a time-varying user distribution model is further considered to explicitly incorporate temporal evolution of the large-scale user geometry and short-term user mobility correlations. In the time-varying model, the channel sample index $s$ is regarded as a discrete-time snapshot taken at $t_s = s \delta T_{s}$, where $\delta T_{s}$ denotes the short-timescale sampling interval corresponding to the update period of the PRA states. The mechanically adjustable array positions are updated on a long timescale and remain unchanged over $\delta T_{l}/ \delta T_{s}$ consecutive snapshots, where $\delta T_{l}$ denotes the long-timescale update period.

Users are generated within a 3D spherical annulus $\mathcal{A}$ with radial distances ranging from $50$~m to $120$~m from the BS center. Three hotspot regions $\mathcal{A}_1$, $\mathcal{A}_2$, and $\mathcal{A}_3$ are spheres with a fixed radius, and the remaining region is denoted by $\mathcal{A}_0$. The hotspot centers are initialized as $\mathbf{c}_i(0)=\bar{\mathbf{c}}_i$ for $i=1,2,3$, where $\bar{\mathbf{c}}_1=(50,-30,-40)$~m, $\bar{\mathbf{c}}_2=(-50,-10,50)$~m, and $\bar{\mathbf{c}}_3=(-10,60,20)$~m. The center of hotspot $\mathcal{A}_i$ at snapshot $s$ is denoted by $\mathbf{c}_i(s)$ and follows a first-order Gauss-Markov process
\begin{equation}
\mathbf{c}_i(s)=\bar{\mathbf{c}}_i+\rho_c\big(\mathbf{c}_i(s\!-\!1)-\bar{\mathbf{c}}_i\big)+\boldsymbol{\epsilon}_i(s),
\quad \boldsymbol{\epsilon}_i(s)\sim\mathcal{N}(\mathbf{0},\sigma_c^2\mathbf{I}_3),
\end{equation}
where $\rho_c\in(0,1)$ controls the temporal smoothness of the center drift. In the simulations, $\sigma_c = 0.05$ determines the drift scale per snapshot.

For each $s\in\{1,\ldots,S\}$, let $K_i(s)$ denote the number of users in $\mathcal{A}_i$ for $i\in\{0,1,2,3\}$. The long-term mean user count is $\mu_i=\mathbb{E}[K_i(s)]$, and the instantaneous counts follow the Poisson distribution.
To introduce short-term correlation across snapshots while keeping $\mathbb{E}[K_i(s)]=\mu_i$ unchanged, from snapshot $(s\!-\!1)$ to $s$, each user in $\mathcal{A}_i$ independently leaves with probability $(1-\rho_i)$ and remains with probability $\rho_i$, which yields the number of surviving users
\begin{equation}
S_i(s)\sim\mathrm{Binomial}\!\left(K_i(s\!-\!1),\,\rho_i\right),
\end{equation}
and the number of newly arrived users is generated as
\begin{equation}
B_i(s)\sim\mathrm{Poisson}\!\left((1-\rho_i)\mu_i\right),
\end{equation}
so that $K_i(s)=S_i(s)+B_i(s)$ holds for all $i\in\{0,1,2,3\}$.

Conditioned on the hotspot center $\mathbf{c}_i(s)$, the locations of users inside $\mathcal{A}_i$ are generated with short-term spatial continuity. For a surviving user $u$ in hotspot $\mathcal{A}_i$, let $\mathbf{p}_{u}(s)$ denote its position at snapshot $s$. Define the relative offset $\mathbf{r}_u(s)=\mathbf{p}_u(s)-\mathbf{c}_i(s)$ and update it by
\begin{equation}
\mathbf{r}_u(s)=\rho_p\mathbf{r}_u(s\!-\!1)+\boldsymbol{\eta}_u(s),\quad
\boldsymbol{\eta}_u(s)\sim\mathcal{N}(\mathbf{0},\sigma_p^2\mathbf{I}_3),
\end{equation}
where $\rho_p\in(0,1)$ controls the cross-snapshot correlation of user motion inside the hotspot. In the simulations, $\rho_p =0.95$ and $\sigma_p = 0.6$ are used. The user position is given by $\mathbf{p}_u(s)=\mathbf{c}_i(s)+\mathbf{r}_u(s)$. Newly arrived users in each region are drawn uniformly.

\begin{figure*}[!t]
		\centering
		\setlength{\abovecaptionskip}{+4mm}
		\setlength{\belowcaptionskip}{+1mm}
		\subfigure[User spatial distribution (channel sample $s=1$).]
        {\includegraphics[width=2in, height = 1.5in]{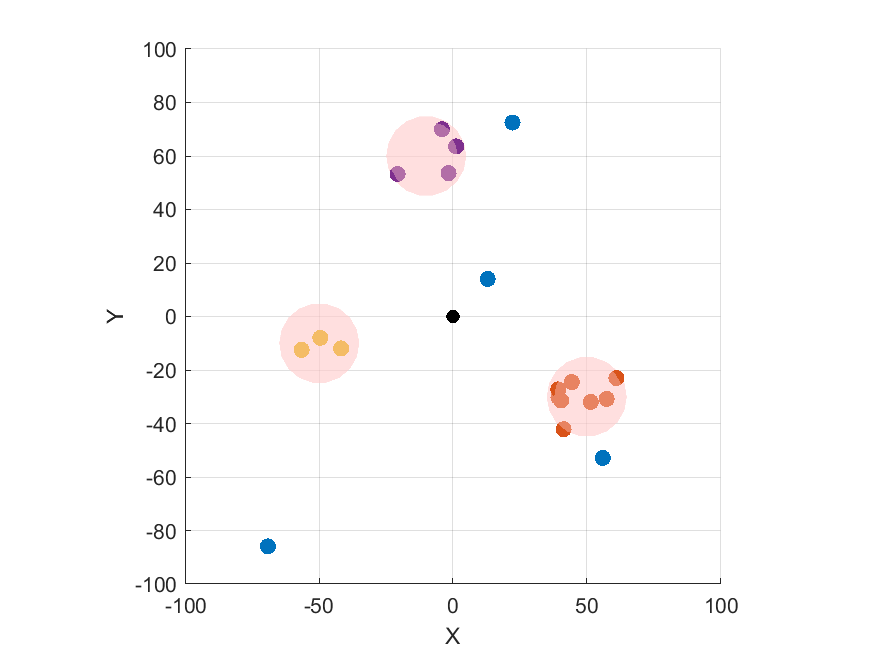}}
        \subfigure[User spatial distribution (channel sample  $s=2$)]
        {\includegraphics[width=2in, height = 1.5in]{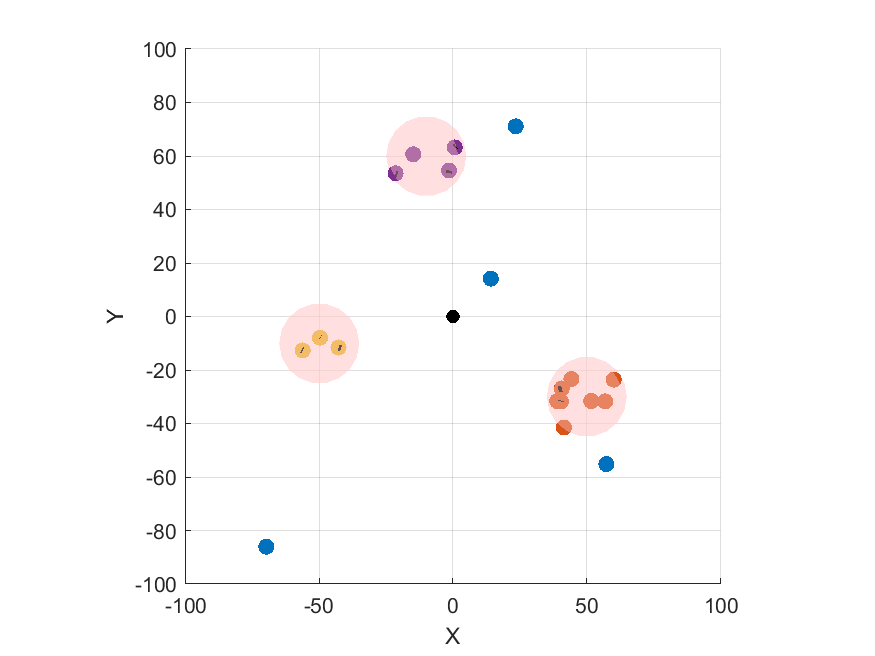}}
        \subfigure[User spatial distribution (channel sample  $s=3$)]
        {\includegraphics[width=2in, height = 1.5in]{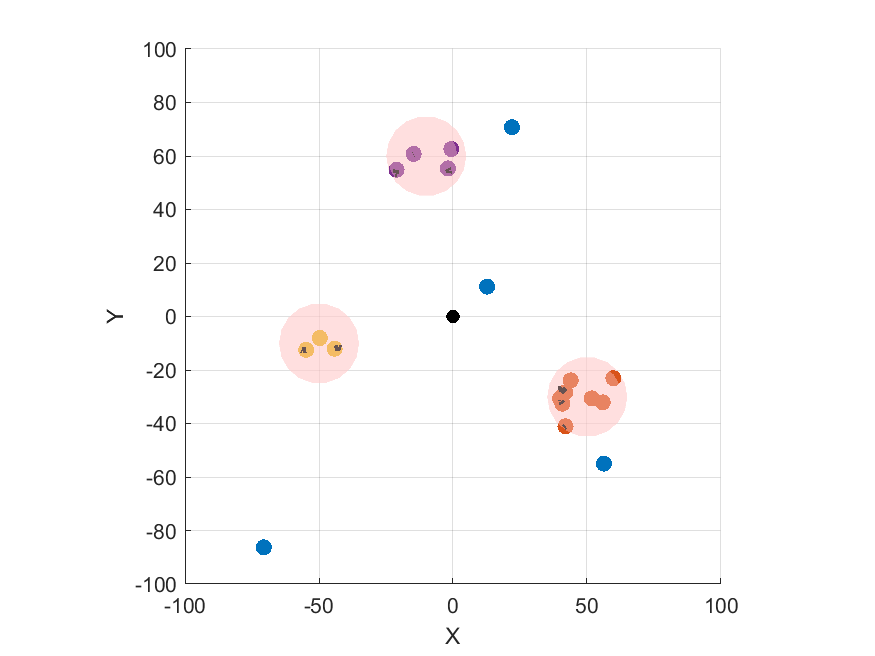}}
        \subfigure[User spatial distribution (channel sample  $s=4$).]
        {\includegraphics[width=2in, height = 1.5in]{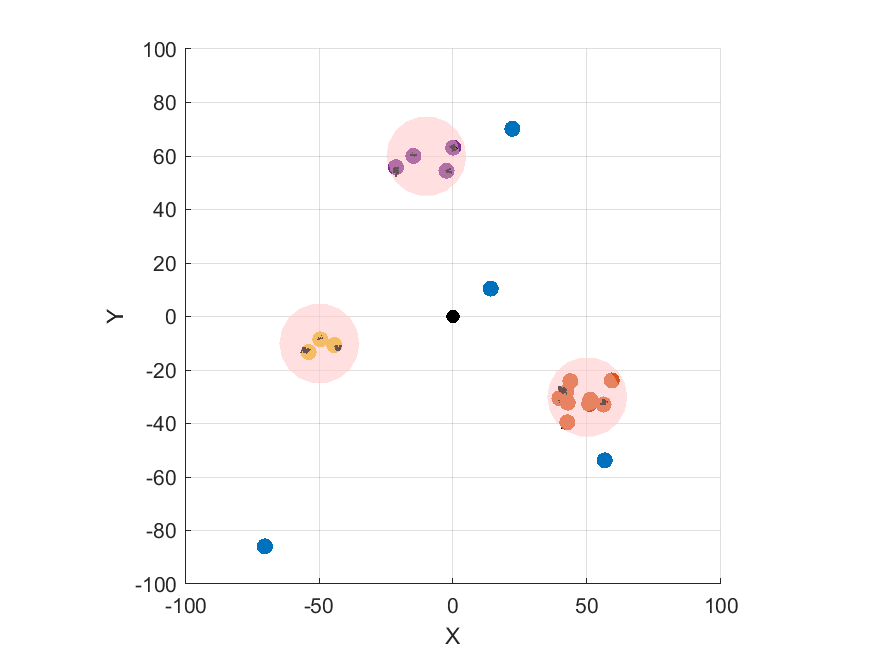}}
        		\subfigure[User spatial distribution (channel sample $s=5$)]
        {\includegraphics[width=2in, height = 1.5in]{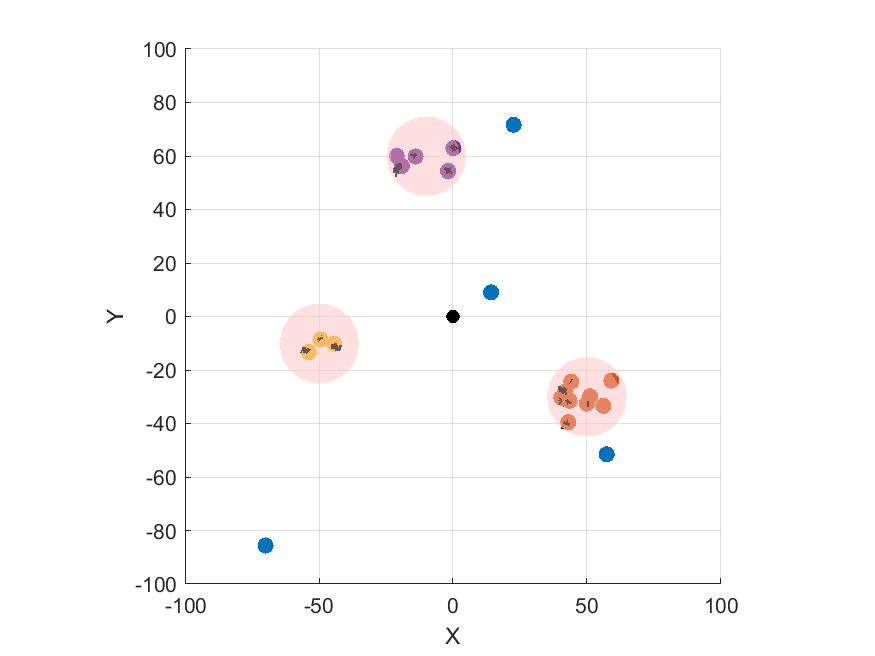}}
                		\subfigure[User spatial distribution (channel sample $s=10$, with trajectory plotted).]
        {\includegraphics[width=2.1in, height = 1.5in]{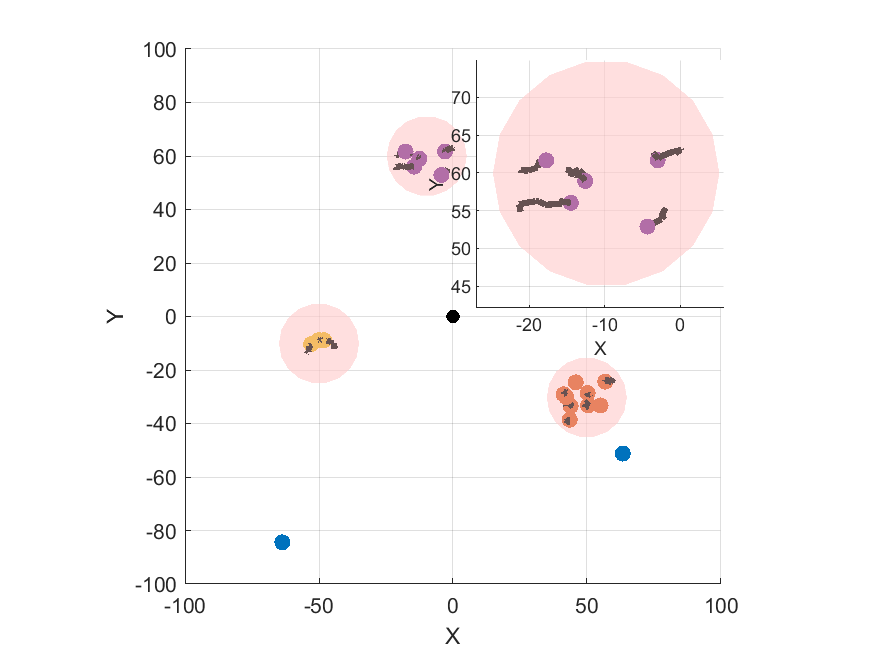}}
		\caption{User spatial distribution from top view ($\rho_c = 0.99$, $\{\rho_i\}_{i=0,1,2,3}=0.98$, $\eta = 0.15$).}
\label{fig:tv_users}
\end{figure*}

\begin{figure}[!t]
\centering
\setlength{\abovecaptionskip}{+2mm}
\setlength{\belowcaptionskip}{-1mm}
\subfigure[Time-varying user distribution.]{\includegraphics[width=1.55in, height=1.35in]{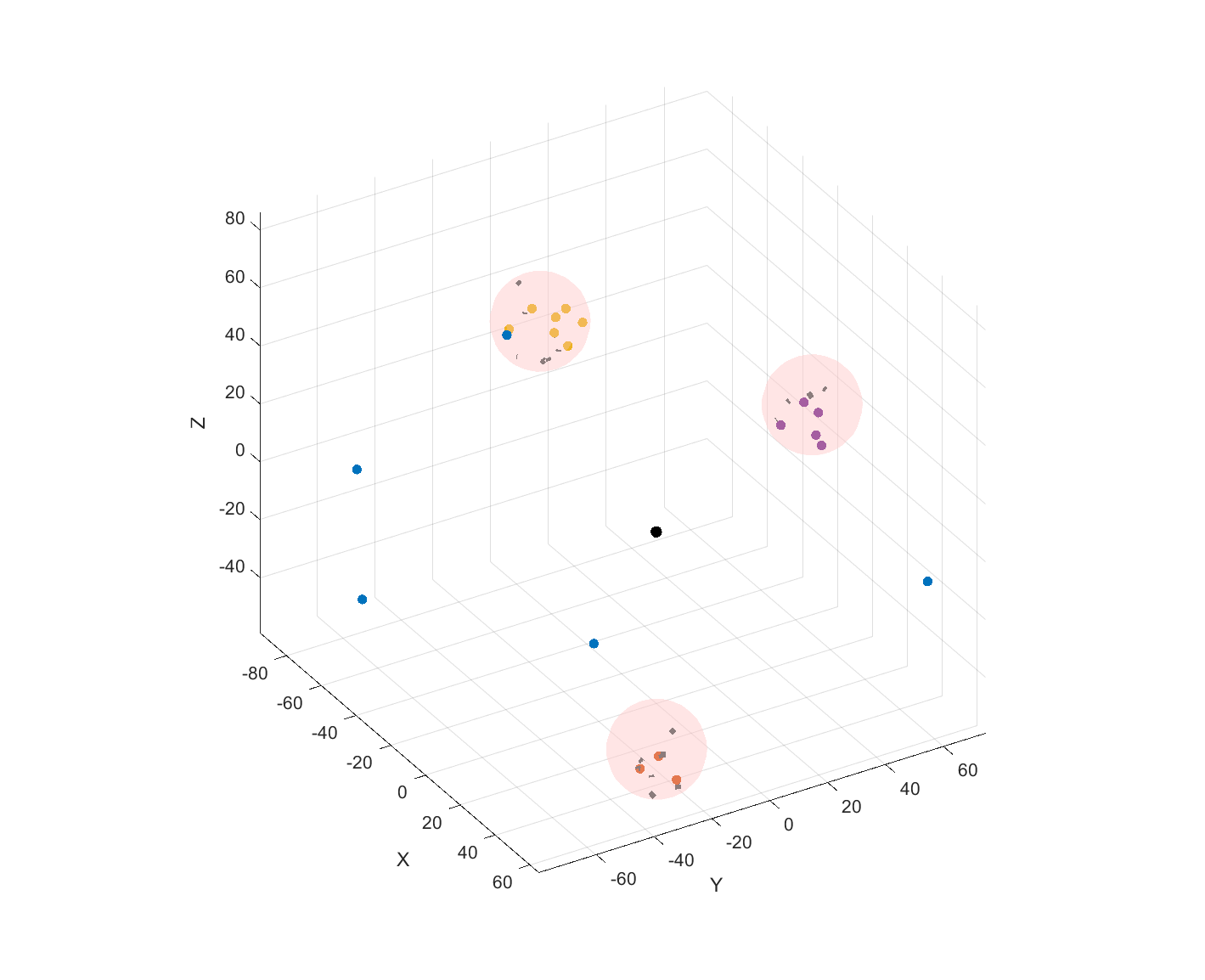}}
\subfigure[Achievable average sum rate versus $\eta$.]{\includegraphics[width=1.55in, height=1.35in]{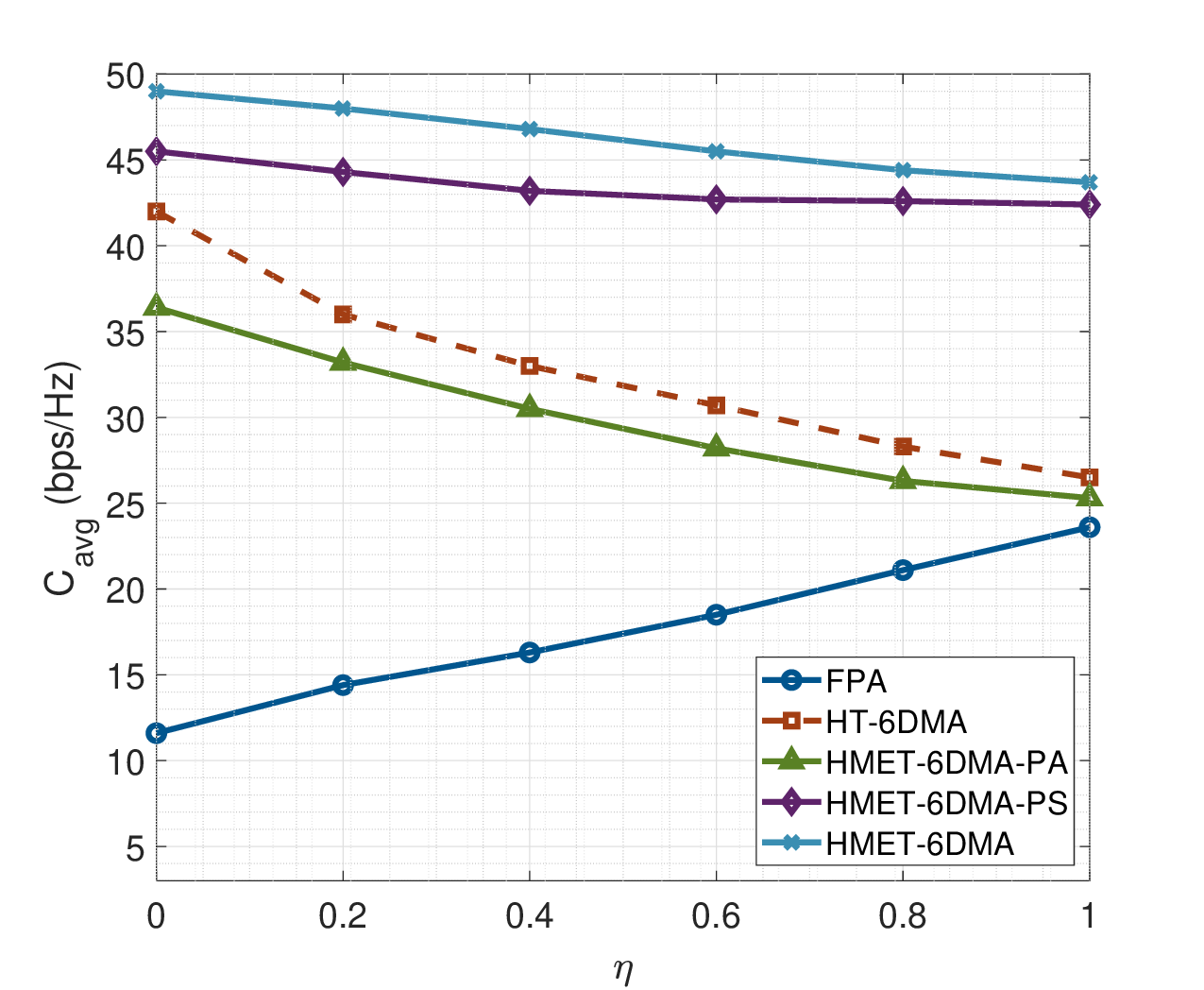}}
\caption{Optimization results under time-varying user distribution.}
\vspace{-0.5cm}
\label{fig:tv_perf}
\end{figure}

Fig.~\ref{fig:tv_users} illustrates representative snapshots of the time-varying user distribution, where the hotspot centers drift mildly and user locations exhibit short-term continuity. Fig.~\ref{fig:tv_perf} compares the achievable average sum rate under the time-varying user distribution. It is observed that the proposed HMET-6DMA still achieves a significant performance gain over all benchmark schemes, which demonstrates the robustness of the proposed hybrid architecture in dynamic environments.

\vspace{-0.2cm}
\section{Conclusion}
\vspace{-0.1cm}
This letter proposes a new HMET-6DMA BS architecture. Each array performs mechanical position adjustment in the long timescale based on the user distribution, while array antennas conduct radiation pattern switching in the short timescale based on the instantaneous user location. To maximize the average sum rate, a two-timescale optimization framework is proposed to jointly optimize the positions of antenna arrays and their antenna radiation pattern selection, by exploiting both the wide-coverage but slow mechanical control and the fast but limited-range electronic control. Simulation results show that the proposed HMET-6DMA BS achieves significant sum-rate gains over various benchmarks, especially when the user distribution exhibits hot-spot clustering.

\end{document}